\documentclass[12pt]{article}

\usepackage{a4wide}
\usepackage[T1]{fontenc}
\usepackage{graphicx}
\usepackage{color}
\usepackage{times} 
\usepackage{amssymb}
\usepackage{cite} 

\newcommand{\gtwo}{\ensuremath{g^{(2)}(0)}}
\newcommand{\nlin}{\ensuremath{n_{l}}}

\title{Atom chips and one-dimensional Bose gases}
\author{I. Bouchoule$^1$, N.J. van Druten$^2$, C. I. Westbrook$^1$\\[5mm]
1: Laboratoire Charles-Fabry, CNRS UMR 8501,  Institut d'Optique, Palaiseau, France\\
2: Van der Waals-Zeeman Instituut, Universiteit van Amsterdam,
 The Netherlands
}

\begin{document}
\maketitle
\tableofcontents

\section{Introduction}
As this volume indicates, the technology of atom chips is currently enjoying
great success for a large variety of experiments on degenerate quantum gases. 
Because of their geometry and their
ability to create highly confining potentials, they are
particularly well adapted to realizing one dimensional (1D) situations
\cite{Rei02,FolKruSch02,ForZim07,WilHofLes06,SchHofAnd05,TreEstWes06,EstTreSch06,AmeEsWic08,epjd-si-ac}.
This characteristic has contributed to a revival of interest in the study
of 1D Bose gases with repulsive interactions, a system which
provides a vivid example of an exactly solvable quantum many-body system
 \cite{LieLin63,Lie63,YanYan69}. 
The quantum many-body eigenstates \cite{LieLin63,Lie63} and
thermodynamics \cite{YanYan69} can be calculated without resorting to
approximations.
In addition, the 1D Bose gas shows a remarkably
rich variety of physical regimes (see Fig.~\ref{fig.domaines})
that are very different both from those found  in 2D and in 3D. 
One dramatic example of the difference is the tendency for a
1D Bose gas to become more
{\em strongly} interacting as its density {\em decreases} \cite{LieLin63}.
Finally, and in a more practical vein, a good
understanding of its behavior is relevant for 
guided-wave atom lasers \cite{GueRioGae06} and 
trapped-atom interferometry \cite{HofLesFis07a}.
Because of the effects of interactions, the analogy to the manipulation
of light in single mode fibers needs to be examined carefully.

An atom chip is not the only means of producing a 1D Bose gas. 
Optical trapping has been used
to generate  similarly  elongated trap geometries.
In particular, a 2D optical lattice can be used to generate a 2D array of 1D traps
\cite{KinWenWei04,KinWenWei05,LabHarHuc04,ParWidMur04,MorStoKoh03}.
Because of the massively parallel nature
of this system, it is possible to work with only a few atoms per tube, 
and still get a sizeable signal per experimental cycle. 
Thus, the strongly
interacting regime alluded to above can be reached.
This regime has yet to be reached with an atom chip. 
But as we will show here, a key feature of atom chips is that they produce
individual samples in which one does not intrinsically average over many realizations.
Fluctuation phenomena are therefore readily accessible, an aspect which we
will treat later in this chapter.

In the following we first give
an introduction to the various regimes of the 
homogeneous 1D Bose gas, with particular emphasis on the behavior
of the density profiles and the density fluctuations in the context of approximate models.
Then we will discuss the exact solution and how it differs from the approximations. 
Next, we discuss some of the important issues involved in
realizing 1D gases in a 3D trap. Finally, we describe a series
of experiments performed in Orsay and Amsterdam 
using atom chips to explore and illustrate features of the
1D Bose gas.

\section[Regimes of 1D gases]{Regimes of one-dimensional gases}
\label{sec.theory}
First, we review some theoretical results concerning 
the one-dimensional Bose gas with repulsive interactions. 
Most of these results are derived in 
Refs.~\cite{LieLin63,YanYan69,Ols98,PetShlWal00,KheGanDru03,KheGanDru05,Pop83}.
Here we will concentrate on intuitive arguments, and the reader is referred to
the above references for more careful demonstrations. 
The system is described by the Hamiltonian
\begin{equation}
H=-\frac{\hbar^2}{2m}\int dz \psi^+ \frac{\partial ^2}{\partial_z^2}\psi
+\frac{g}{2}\int dz \psi^+\psi^+\psi\psi,
\label{eq.ham}
\end{equation}
where $\psi$ is the field operator in second quantization, and $g$ is
the coupling constant characterizing the interactions between particles. 
From this coupling constant, one can deduce an intrinsic length scale related 
to the interactions, 
\begin{equation}
l_g=\frac{ \hbar^2}{mg},
\label{eq.lg}
\end{equation}
as well as an energy scale: 
\begin{equation}
E_g=\frac{ mg^2}{2\hbar^2}=\frac{\hbar^2}{2ml^2_g}.
\label{eq.Eg}
\end{equation}

In thermal equilibrium, the gas is described by
the temperature $T$ and the linear atomic density $n$.
Rescaling these two quantities by the intrinsic scales introduced above, 
and setting Boltzmann's constant equal to unity (i.e., measuring
 temperature in units of energy)  
we find that the properties of the gas
are functions of the dimensionless quantities
\begin{equation}
t=\frac{T}{E_g},
\label{eq.t}
\end{equation}
and 
\begin{equation}
\gamma=\frac{mg}{\hbar^2 n}=\frac{1}{n l_g},
\label{eq.gamma}
\end{equation}
the latter being the famous Lieb-Liniger parameter \cite{LieLin63}.

\begin{figure}[t]
 \centerline{\includegraphics{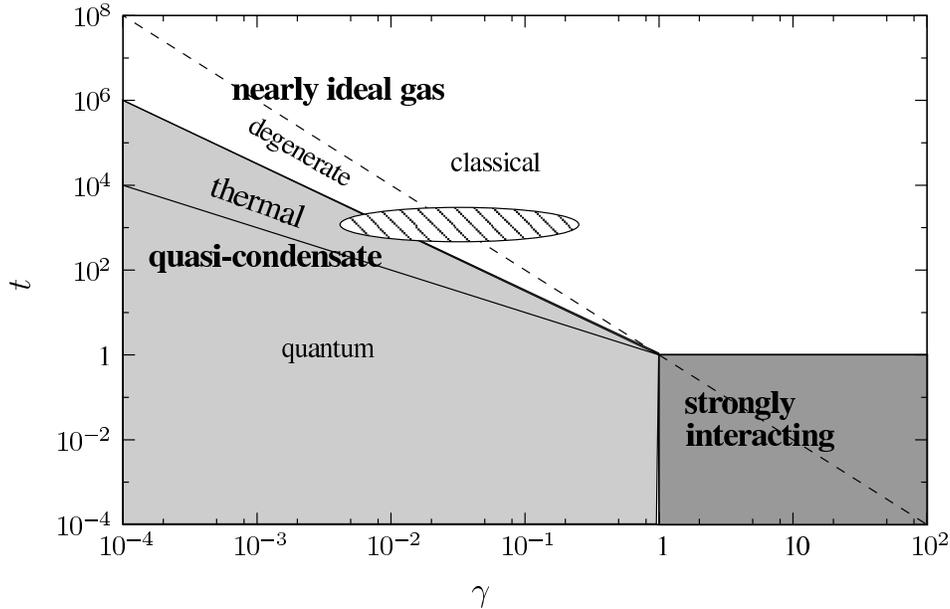}}
\caption{\it Physical regimes of a 1D Bose gas with repulsive contact 
interactions in the parameter space $(\gamma,t)$, adapted
from~\cite{KheGanDru03}. 
The dashed diagonal line separates the degenerate and nondegenerate gases. 
The strongly interacting regime is shown in dark grey.
The weakly interacting regime is 
divided into the nearly ideal gas regime (also called decoherent regime)
shown in white and the 
quasicondensate regime shown in light grey.
Note that the nearly ideal gas can be degenerate. 
The quasicondensate regime is divided into 
the  thermal and quantum regimes. 
The lines represent smooth  (and often wide) 
crossovers rather than phase transitions.
The crossovers are given in 
Eqs.~(\ref{eq.strongweak}), (\ref{eq.frontTI}), (\ref{eq.degen}),
(\ref{eq.frontncohom}) and (\ref{eq.qbecthqu}).
The dashed area shows the parameter space investigated in 
the experiments presented in this chapter.
}
\label{fig.domaines}
\end{figure}

It 
is useful to also introduce two other relevant scales, namely the thermal 
de Broglie wavelength,
\begin{equation}
\lambda_{dB}=\hbar\sqrt{\frac{2\pi}{mT}}=l_g\sqrt{\frac{4\pi}{t}},
\end{equation}
and the quantum degeneracy temperature 
\begin{equation}
T_d=\frac{\hbar^2 n^2}{2m}=\frac{E_g}{\gamma^2}.
\label{eq.Td}
\end{equation}
In the above $(t,\gamma$) parametrization, quantum
degeneracy  ($T\approx T_d$, or equivalently $n\lambda_{dB}\approx 1$)
is reached around
\begin{equation}
t\approx \frac{1}{\gamma^2}.
\label{eq.degen}
\end{equation}

The thermal 
equilibrium for the hamiltonian of Eq.~(\ref{eq.ham}) has 
been extensively studied theoretically~\cite{YanYan69,KheGanDru03}.
Without going into great detail however, we can present some important
features of this system. 
Several regimes may be identified in the parameter space $(\gamma, t)$, 
as sketched in Fig.~\ref{fig.domaines}.
We begin by noting that the region $\gamma \gg 1$, $t\ll 1$ (dark grey area) 
defines a strongly interacting regime that occurs at 
low density and low temperature, often
referred to as the Tonks-Girardeau gas \cite{Ton36,Gir60,Ols98}.

In the weakly interacting regime, $\gamma<1$, several sub-regimes 
are identified. 
These are the regimes which to date have been 
accessible in atom chip experiments, and
we shall elaborate further on their nature in the discussion below. 
The two main regimes are  the nearly ideal gas regime (white area) 
and the 
quasi-condensate regime (light grey area). 
Each one permits an approximate description that we 
present later in this section 
and which allows the identification of sub-regimes.
For the moment we simply wish to emphasize that no phase transition occurs 
in the 1D Bose gas and that 
all the boundaries represent smooth (and often broad) 
crossovers in behavior. 

\subsection{Strongly versus weakly interacting regimes}
\label{sec.weakvsstrong}
\begin{figure}
\centerline{\includegraphics{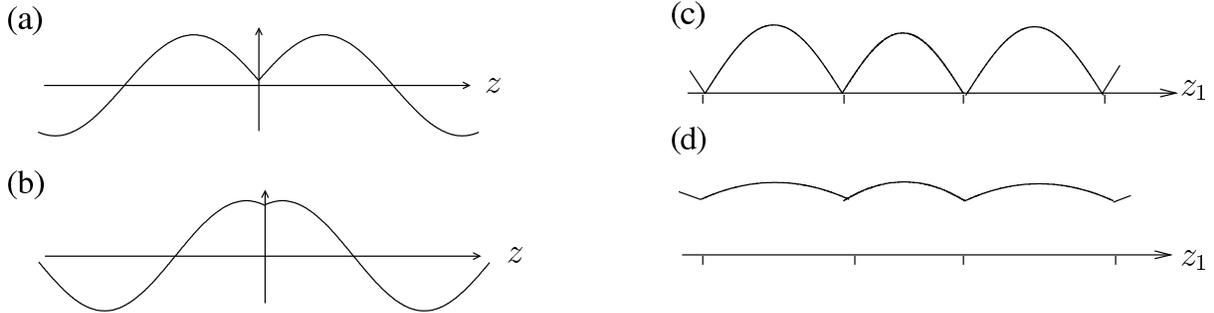}}
\caption{\it 
Strong interaction versus weak interaction regime. 
We show the  wave function  in the center-of-mass frame of two atoms 
for (a) strong interactions,  
scattering energy $E$ much smaller than $E_g=mg^2/2\hbar^2$
and (b) weak interactions, $E$ much larger than $E_g$.
We also plot the  wave function $\psi(z_1,z_2,z_3,...)$
for given positions of $z_2,z_3,...$ in (c) the strongly interacting regime
and (d) the weakly interacting regime.  
}
\label{fig.regimes1D}
\end{figure}

We first comment on the distinction between strong and weak interactions. 
 Following the approach of~Ref.~\cite{Ols98},
we study the scattering 
wave function of two atoms interacting via the potential 
$g\delta(z_1-z_2)$, where $z_1$ and $z_2$ are the position 
of the two atoms. For this, we consider the wave function $\psi$ 
 in the 
center-of-mass frame, with reduced mass $m/2$ and subject to 
the potential $g\delta(z)$. The effect of the potential 
is described by the continuity condition
\begin{equation}
\frac{\partial }{\partial_z}\psi(0_+)-
\frac{\partial }{\partial_z}\psi(0_-)=\frac{mg}{2\hbar^2}\psi(0)
\label{eq.continuity}
\end{equation}
 where $0_+$ ($0_-$) denotes the limit when $z$ goes to zero
through positive (negative) values.
Let us consider the scattering solution for an energy $E=\hbar^2 k^2/m$.
Since we consider bosons, we look for even wave functions of the form
$\cos(k|z|+\phi)$. The continuity conditions give 
$\phi$ and thus the value $\psi(0)$. 
We find then that the energy $E_g$ given by Eq.~(\ref{eq.Eg}) 
is the relevant energy scale and that
for $E\ll E_g$, $\psi(0)$ is close to 
zero, while, for $E\gg E_g$, $\psi(0)$ is close to one, 
as illustrated in Fig.~\ref{fig.regimes1D}.

The above results hold for a gas of particles since the 
continuity relation~(\ref{eq.continuity}) holds for the many-body 
wavefunction 
when two atoms are close to the same place. Thus, as 
long as  the typical energy 
of the particles is much lower than  $E_g$, 
the many-body wavefunction vanishes when two particles are 
at the same position:
the gas is then in the strongly 
interacting, or  Tonks-Girardeau regime. The vanishing of the wave function 
when two particles are at the same place mimics the 
Pauli exclusion principle and the gas acquires some similarities 
with a gas of non interacting fermions. 
 More precisely, 
in this strong interaction regime, 
the available wave functions
of the many body problem are, 
up to a symmetrization factor, the wave functions 
of an ideal Fermi gas~\cite{Gir60}.
Since the 
wave function vanishes when two atoms are at the same place,
the energy of the system is purely kinetic energy and 
 the eigen energies are those of the Fermi system. Thus the
1D strongly interacting Bose gas and the ideal 1D Fermi gas   
share the same energy spectrum. This implies in particular 
that all thermodynamic quantities are identical for both systems.

To identify the parameter space of the 
strongly interacting regime, we suppose the gas to be 
strongly interacting and then require that the typical energy of the atoms be smaller than $E_g$.
To estimate the typical energy per atom, we use the Bose-Fermi 
mapping presented above.  
If the gas is degenerate, the temperature is smaller than the degeneracy temperature
$T_d$, Eq.~(\ref{eq.Td}), and $T_d$ corresponds to the "Fermi" energy of the atoms. 
The typical atom energy is therefore $T_d$ and it is of order $E_g$ if
\begin{equation}
\gamma\simeq 1.
\label{eq.strongweak}
\end{equation}
The strongly interacting regime thus requires $\gamma \gg 1$.
If the gas is non degenerate, the typical energy of the equivalent Fermi gas is $T$ and
interactions become strong when $T=E_g$ or 
\begin{equation}
t\simeq 1.
\label{eq.frontTI}
\end{equation}
We then find that the gas is strongly interacting for $t\ll 1$.

The condition (\ref{eq.strongweak}) is often derived using 
the following alternative argument, valid at zero temperature.
At zero temperature, there are two extremes for the possible 
solutions for the wave function $\psi(z_1,z_2,...)$. 
As seen in Fig.~\ref{fig.regimes1D}, 
either the wave function vanishes when two 
atoms are at the same place, or the wave function is almost 
uniform, corresponding to the strongly and weakly interacting configurations respectively.
In the weakly interacting configuration, the kinetic 
energy is negligible and the interaction energy per particle, of the order of $gn$, determines
the total energy. 
In the strongly interacting configuration, on the other hand, 
the interaction energy vanishes while the typical kinetic energy per 
particle is $\hbar^2 n^2/m$. Comparing these two energies, we find that 
the strongly interacting configuration is favorable only for
$\gamma > 1$.

\subsection{Nearly ideal gas regime}
\label{sec.idealgas}
 At sufficiently high temperatures, 
interactions between atoms have  little effect and the 
gas is well described by an ideal Bose gas. 
In Ref.~\cite{KheGanDru03}, this regime was referred to as
the "decoherent regime"; We will call it 
the  (nearly) ideal Bose gas regime. 
A 1D ideal Bose gas at thermal equilibrium is well described
using the grand canonical ensemble, introducing the chemical 
potential $\mu$.
All properties of the gas are calculated using the Boltzmann 
law which states that, for a given one-particle state of momentum 
$\hbar k$, the probability to find $N$ atoms in this state 
is proportional to $e^{-(\hbar^2k ^2/(2m)-\mu)N/T}$; note that $\mu<0$ in
this description.
In the following, we use a quantization box of size $L$  (tending to
infinity in the thermodynamic limit)
and periodic boundary conditions so that the available 
states are the momentum states with momentum $k=2\pi j/L$ 
where $j$ is an  integer.

Let us first consider the linear gas density.
 From the Boltzmann law, we find that the mean population
is the Bose distribution
\begin{equation}
\langle n_k\rangle =\frac{1}{e^{(\hbar^2k^2/(2m)-\mu)/T}-1}.
\label{eq.bosedistr}
\end{equation} 
 The atom number, and thus the linear density, is obtained by 
summing the population over the  states and one finds 
\begin{equation}
n= \frac{1}{\lambda_{dB}}g_{1/2}(e^{\mu/T}),
\label{eq.stateid}
\end{equation}
 where 
$g_{1/2}(x)$ is one of the Bose functions 
\begin{equation}
g_n(x)=\sum_{l=1}^{\infty}\frac{x^l}{l^n},
\end{equation}
also known as the polylogarithmic functions \cite{Hua87,Lew81}.
Unlike in 3D systems, where the excited-state density is given by
$\rho_e=g_{3/2}(e^{\mu/T})/\lambda_{dB}^3$ 
in this approach \cite{Hua87}, 
no saturation of the excited states occurs (the function $g_{1/2}$  diverges 
as  $\sqrt{-\pi T/\mu}$ as $\mu\rightarrow 0$ from below, whereas $g_{3/2}(1)=2.612$ is finite):
in the thermodynamic limit no Bose-Einstein condensation is 
expected and the gas is well described by a thermal 
gas at any density.

Two asymptotic regimes may be identified: 
the non degenerate regime for which $-\mu\gg T$ and $\hbar^2n^2/m\ll T$
and the degenerate regime for which $-\mu\ll T$ and $\hbar^2n^2/m\gg T$. 
In the non degenerate regime, the linear density is well approximated 
by the 
Maxwell-Boltzmann formula
\begin{equation}
n= \frac{1}{\lambda_{dB}}e^{\mu/T},
\label{eq.stateMB}
\end{equation}
In this regime $n \lambda_{dB}$ is much smaller than unity.
In the degenerate regime, the states of energy much 
smaller than $T$ are highly occupied and the linear 
density is given by 
\begin{equation}
n=\frac{T}{\hbar}\sqrt{\frac{m}{-2\mu}}
\label{eq.mudegenerate}
\end{equation}
This density is much larger than $1/\lambda_{dB}$, {\it i.e. 
$n\lambda_{dB}\gg 1$}.

As we will discuss in the experimental section, 
fluctuations are also very important for characterizing the gas.
 It is thus instructive to consider the correlation functions.  
The  normalized one body correlation function is 
$g^{(1)}(z)=\langle \psi^+(0)\psi(z)\rangle/n$, where $\psi$ is the  
field operator in the second quantization picture. 
Using the expansion of the field operator in the plane wave basis 
$\psi(z)=\sum_k a_ke^{-ikz}/\sqrt{L}$ where $a_k$ is 
the annihilation operator for the mode $k$, 
we find $g^{(1)}(z)=\sum_k  \langle n_k\rangle e^{-ikz}/(Ln)$.
Here $n_k=a_k^+a_k$ is the atom number operator for the mode $k$.
 Simple 
analytical expressions are found in the nondegenerate and highly degenerate limits. 
In the non degenerate limit ($-\mu\gg T$ 
or, equivalently $n\ll 1/\lambda_{dB}$), we find 
\begin{equation}
g^{(1)}(z)\simeq e^{-\frac{z^2}{4\pi\lambda_{dB}^2}}.
\label{eq.g1gasiddegenerate}
\end{equation}
As the gas becomes more degenerate, the correlation length 
increases and, in the degenerate regime ($-\mu\ll T$ 
or, equivalently $n\gg 1/\lambda_{dB}$), we find 
\begin{equation}
g^{(1)}(z)\simeq e^{-\frac{mT z}{n\hbar^2}}=e^{-\frac{2\pi z}{n\lambda_{dB}^2}}.
\label{eq.g1iddegene}
\end{equation}
In this regime the correlation length, 
about $n\lambda_{dB}^2$, is much larger than the de Broglie wavelength
(and the mean interparticle distance $1/n$)  since
$\lambda_{dB}\gg 1/n$.

Next we consider the normalized density-density or two body correlation function 
\begin{equation}
g^{(2)}(z)=\langle \psi^+(z)\psi^+(0)\psi(0)\psi(z)\rangle/n^2. 
\end{equation}
This function is proportional to the probability of finding an atom at position
$z$ {\em and} at position $z=0$.
It  is given by
\begin{equation} 
n^2g^{(2)}(z)=\sum_{k_1k_2k_3k_4} \langle a_{k_1}^+a_{k_2}^+a_{k_3}a_{k_4}\rangle
e^{ik_1z}e^{-ik_4z}/{L^2 }.
\end{equation}
Using Bose commutation relations and 
the fact that, since atoms do not interact, 
different momentum state populations are uncorrelated,
the sum simplifies to:
\begin{equation} 
n^2 g^{(2)}(z)=\sum_{k_1\neq k_2} \langle n_{k_1}\rangle \langle n_{k_2}\rangle (1+e^{i(k_1-k_2)z})/L^2
+\sum_k \langle a_{k}^+a_{k}^+a_{k}a_{k}\rangle/L^2.
\label{eq.g2id}
\end{equation}
In the last term, the commutation relations give:
 $\langle a_{k}^+a_{k}^+a_{k}a_{k}\rangle = 
 \langle n_k^2 \rangle-\langle n_k\rangle$, and in thermal equilibrium one has: 
 \begin{equation}
\langle n_k^2 \rangle=\langle n_k\rangle + 2\langle n_k\rangle^2.
\label{eq.bunching1etat}
\end{equation}
Therefore we find:
\begin{equation}
g^{(2)}(z)=1+|g^{(1)}(z)|^2,
\label{eq.g2wick}
\end{equation}
a result which one can also obtain directly from Wick's theorem \cite{Wic50}.
Equation~(\ref{eq.g2wick}) means that the probability of finding atoms 
within less than a correlation length in a thermal Bose gas 
is twice that of finding two atoms far apart. 
This phenomenon is often referred to
as "bunching" and has been observed in cold atoms in 
several experiments \cite{YasShi96,FolGerWid05,HBTHe}.
Bunching is closely related to density fluctuations.
As one can see from Eq.~(\ref{eq.bunching1etat}), in a thermal gas, fluctuations in the
occupation of a single quantum state, 
$ \delta {n_k} ^2  = \langle n_k^2 \rangle-\langle n_k\rangle^2$,
show a "shot noise" term, $\langle n_k\rangle$ and an "excess noise" term, $\langle n_k\rangle^2$.
The density fluctuation experiment described later in this chapter has 
demonstrated this behavior. 

{\it Validity of the ideal gas treatment.}
The two body correlation function has been used to characterize the crossover between the 
ideal gas and quasi-condensate regimes \cite{KheGanDru03}.
When interactions become important, they impose an energy cost on density
fluctuations and the latter tend to smooth out. 
This amounts to a reduction in the value of $g^{(2)}(0)$. 
In the quasi-condensate regime which we discuss in the next section, 
the bunching effect is absent and $g^{(2)}(0)$ is close to unity. 
The ideal Bose gas description fails when the typical interaction energy 
per particle
$gn$ is not negligible compared to $-\mu$. 
Using  Eq.~(\ref{eq.mudegenerate}) 
one finds that the ideal Bose gas description fails when the temperature 
is no longer much smaller than the crossover temperature, which we define as 
\begin{equation}
T_{co}\simeq T_d\sqrt{\gamma}.
\label{eq.ncohom}
\end{equation}
Using the reduced dimensionless temperature $t=T/E_g$, 
this can be written as
\begin{equation}
t_{co}\simeq \frac{1}{\gamma^{3/2}}.
\label{eq.frontncohom}
\end{equation}
This line separates the nearly ideal gas regime from the 
quasi-condensate regime in Fig.~\ref{fig.domaines}.
Note that, in terms of chemical potential, the domain of validity of the 
ideal gas model is $-\mu\gg \mu_{co}$ where
we define the crossover chemical potential as
\begin{equation}
\mu_{co}=\frac{T}{t^{1/3}}.
\label{eq.muco}
\end{equation}
In making this estimate, we have assumed that the gas is degenerate at the crossover. 
From Eq.~(\ref{eq.ncohom}), one can see that if one is in the weakly interacting regime
($\gamma \ll 1$) this assumption is indeed true. 
The experiments described below confirm that one can observe the effects of 
degeneracy before the onset of the reduction of density fluctuations.  

A precursor of the reduction of density fluctations is shown by 
a perturbative calculation valid in the nearly ideal gas regime which gives, 
to lowest order 
in $g$~\cite{KheGanDru03}, 
\begin{equation}
g^{(2)}(0)\simeq 2- 4(T_{co}/T)^2.
\end{equation}
 To accurately treat the crossover regime however, it is necessary to 
make use of the exact solution to the 1D Bose gas model. 
The exact solution in the crossover regime is discussed in Sec.~\ref{sec.exact}.

The correlation lengths of the gas
are important parameters of the gas that 
will be used in the following to estimate the validity criteria of the 
local density approximation. In the degenerate regime, the correlation
length is $l_c\simeq n \lambda_{dB}^2$ (see Eq.~(\ref{eq.g1iddegene})). 
Using Eq.~(\ref{eq.ncohom}), we find that, close to the crossover, the 
correlation length of the gas is close to the healing length 
\begin{equation}
\xi=\frac{\hbar}{\sqrt{mgn}}.
\label{eq.xi}
\end{equation}

\subsection{Quasi-condensate regime}
\label{sec.quasibec}
On the other side of the crossover, {\it i.e.} for $T\ll T_{co}$, 
the bunching effect is entirely suppressed and the 
$g^{(2)}$ function is close to unity for any $z$. 
This regime is the quasi-condensate regime\footnote{It is also called coherent 
regime since the $g^{(2)}$ function is close to unity, as 
in a coherent state. On the other hand, the first order correlation function
still decays and so the gas is not strictly coherent in this sense.
Within this terminology, the ideal Bose gas regime is called the decoherent 
regime~\cite{KheGanDru03}.}.
 In this section, we present a description of the gas, 
valid in the quasi-condensate regime. This description permits a 
simple estimate of the density fluctuations. 
We thus verify {\it a posteriori} that the quasi-condensate regime 
is obtained for $T\ll T_{co}$. We also give 
a simple calculation of phase fluctuations in the quasi-condensate regime.

In the quasi-condensate regime density fluctuations 
are strongly reduced compared to their value in an ideal Bose gas
where the bunching effect is responsible for density fluctuations 
of the order of $n^2$. 
In other words:
\begin{equation} 
\delta n ^2  \ll n^2
\label{eq.condfluctusmall}
\end{equation} 
In this regime, a suitable description is realized by 
writing the field operator as $\psi=e^{i\theta}\sqrt{n+\delta n}$
where the real number $n$ is the mean density
and  the operator $\delta n$ and the phase operator $\theta$
are conjugate: $[\delta n(z),\theta(z')]=i\delta(z-z')$.
Note that the definition of a local phase operator is subtle and 
the condition Eq.~(\ref{eq.condfluctusmall}) is not well 
defined since, because of shot noise, $\delta n^2$ is expected 
to diverge in a small volume.
A rigorous and  simple approach consists in discretizing the space 
so that in each cell a large number of atoms is present while
the discretisation step is much smaller than the correlation 
length of density and phase fluctuations \cite{Quasibec_Castin}.  

Following this prescription, 
one first minimizes the grand canonical 
Hamiltonian $H-\mu N$ with respect to $n$
to obtain the equation of state 
\begin{equation}
\mu = gn.
\label{eq.muqbec}
\end{equation}
 To second order 
in $\delta n$, this is the correct expression of the chemical potential.
This equality ensures that the Hamiltonian has no linear terms 
in  $\delta n$ and $\nabla \theta$.  
 Linearizing the Heisenberg equations of motion in $\delta n$ 
and $\nabla \theta$, we obtain~\cite{Quasibec_Castin}  
\begin{equation}
\left\{ 
\begin{array}{l} 
\hbar \partial \theta/\partial t=-\frac{1}{2\sqrt{n}}(-\frac{\hbar^2}{2m}\Delta
+2gn )\frac{\delta n}{\sqrt{n}}\\
\hbar \partial \delta n /\partial t=2\sqrt{n}(-\frac{\hbar^2}{2m}\Delta
)\theta \sqrt{n} 
\end{array} 
\right .
\label{eq.hydro} 
\end{equation} 
These equations 
are the so-called hydrodynamic equations. They are derived from
a Hamiltonian quadratic in $\delta n$ and $\nabla \theta$, that can be 
diagonalized using the Bogoliubov procedure~\cite{Quasibec_Castin}. 
It is not the purpose of this 
chapter to detail this calculation and to give exact results within 
this theory. 
We will simply give  arguments that 
enable an estimate of  the 
density fluctuations and of their correlation length. 
This estimate will then be used  to check that $\delta n^2\ll n^2$, 
as assumed in Eq.~(\ref{eq.condfluctusmall}).
We will 
show that this condition is the same as
the condition $T\ll T_{co}$ where $T_{co}$
 given in
Eq.~(\ref{eq.ncohom}).
After that, we will give similar arguments to estimate the phase fluctuations.
Since in the following we will study the gas properties versus the 
chemical potential, it is instructive to rewrite the condition $T\ll T_{co}$ 
in terms of chemical potential. Using Eq.~(\ref{eq.muqbec}), we 
find that the quasi-condensate regime is valid as long as $\mu\gg \mu_{co}$ where 
$\mu_{co}$ is given by Eq.~(\ref{eq.muco}).

\subsubsection{Density fluctuations}
 To estimate the density fluctuations introduced by the excitations, 
it is convenient to divide the excitations in two groups: the 
excitations of low wave vector for which the phase representation is most 
appropriate and the excitations of high wave vector for which a particle 
point of view is most convenient.

 In the following,  we use 
the expansions on sinuso\"idal modes
$\theta=\sum_{k>0}\sqrt{2}(\theta_{ck}\cos(kz)+\theta_{sk}\sin(kz))$ 
and 
$\delta n=\sum_{k>0}\sqrt{2}(\delta n_{ck}\cos(kz)+\delta n_{sk}\sin(kz))$.
Here $\delta n_{jk}$ and $\theta_{jk}$ are conjugate variables 
($[\delta n_{jk},\theta_{j'k'}]=(i/L)\delta_{jj'}\delta_{kk'}$) where 
$j$ stands for $c$ or $s$. 
 For modes of small wave vector $k$, the excitations are 
phonons, or density waves, for which the relative density 
modulation amplitude $\delta n_{jk} /n$ is much smaller than the phase 
modulation amplitude $\theta_{jk}$.
 In this case, 
the local velocity of the gas is given by $\hbar \nabla \theta/m$
and  
the kinetic energy term is simply 
$Ln\hbar^2 k^2 \theta_{jk}^2/(2m)$. The Hamiltonian for this mode 
then reduces to 
\begin{equation}
H_{jk}=L\left (g\delta n_{jk}^2/2 +n \hbar^2 k^2\theta_{jk}^2/(2m)\right ).
\label{eq.Hphonons}
\end{equation}
This hamiltonian could also have been derived from 
the equations of motion given in Eq.~(\ref{eq.hydro}), provided 
that the quantum pressure term $\hbar^2/(2m)\Delta \delta n/n$
is neglected: indeed, for a given wave vector $k$, the 
laplacians in Eq.~(\ref{eq.hydro}) give a factor $k^2$ and 
Eqs.~(\ref{eq.hydro}) are simply the   
equations of motion derived from the Hamiltonian Eq.~(\ref{eq.Hphonons}).
For temperatures much larger than $ng$, the thermal population 
of these phonon modes is large and classical statistics
apply. 
Thus, the mean energy per quadratic degree of freedom
is $T/2$ and we obtain 
\begin{equation}
\langle \delta n_{jk}^2\rangle=T/(Lg). 
\label{eq.deltank}
\end{equation}
and
\begin{equation}
\langle \theta_{jk}^2\rangle=mT/(Lnk^2\hbar^2). 
\label{eq.deltathetak}
\end{equation}
 We can now check the validity of the
assumption $\delta n_k/n\ll\theta_k$: it 
is valid as long as $k\ll \sqrt{mgn}/\hbar$.
Since $k$ values are spaced by $2\pi/L$, there are 
about $L\sqrt{mgn}/(\pi\hbar)$ modes that satisfy this condition. 
Since the contribution of each of these modes 
to the relative density fluctuations is given 
in Eq.~(\ref{eq.deltank}), 
we find that the contribution of these 
low momentum excitations to the relative density fluctuations is of the order of
\begin{equation}
\frac{\langle \delta n^2\rangle_{\rm phonons}}{n^2}
\simeq \frac{T}{n\hbar\sqrt{gn/m}}\simeq \frac{T}{T_d \sqrt{\gamma}}  
\simeq T/T_{co}
\label{eq.fluctunqbec}
\end{equation}

 For wave vectors much larger than $\sqrt{mgn}/\hbar$, the phase-density 
representation is not the 
most appropriate. 
An excitation of wave vector $k\gg \sqrt{mgn}/\hbar$  
corresponds to the presence of an atom of 
momentum $k$, whose wave function 
is $e^{ikz}/\sqrt{L}$ and whose energy is $\hbar^2k^2/(2m)$.
The anihilation operator for this mode is $a_k$ as introduced
in section~\ref{sec.idealgas}.
 For temperatures much larger than $\hbar^2 k^2/m$, the thermal 
population of this mode is large and  classical field theory, 
in which $a_k$ is treated as a c-number, is adequate.
We then find that $a_k$ has a Gaussian distribution which satifies 
$\langle |a_k|^2\rangle =2mT/(\hbar^2 k^2)$. 
  The density fluctuations 
caused by the presence of such high momentum atoms result 
mainly from the interference between 
the atomic field $a_k e^{ikz}/\sqrt{L}$ 
and the atomic field of long wavelength spatial variations, whose 
amplitude is close to $\sqrt{n}$. The density fluctuations are thus  
$\delta n=\sqrt{n}(a_k e^{ikz}+a_k^* e^{-ikz})/\sqrt{L}$. 
We then find that the 
contribution of the mode of wave vector $k$ to density fluctuations 
is $\delta n_k^2=4nmT/(L\hbar^2k^2)$.
Summing the contributions of the modes for all $k>\sqrt{mgn}/\hbar$, 
we obtain an estimate of the density fluctuations
$\langle \delta n^2\rangle_{\rm atoms}$ caused 
by high momentum excitations:
\begin{equation}
\frac{\langle \delta n^2\rangle_{\rm atoms}}{n^2}
\simeq \frac{T}{T_d\sqrt{\gamma}}\simeq\frac{T}{T_{co}}.
\label{eq.fluctunqbecatoms}
\end{equation} 
One also sees from  the above argument that the density fluctuations fall off as
$1/k^2$ above $k=\sqrt{mgn}/\hbar$.
The inverse of this scale gives the 
length scale of density fluctuations and we find that this 
correlation length is the healing length $\xi$ defined in Eq.~(\ref{eq.xi}).

From Eq.~(\ref{eq.fluctunqbec}) and Eq.~(\ref{eq.fluctunqbecatoms}), we 
find that $\delta n^2/n^2\simeq T/T_{co}$. Thus, 
the quasi-condensate treatment is valid as long as 
$T\ll T_{co}$,
 In conclusion, we have shown that $T_{co}$ gives the limit of both the ideal 
gas regime, valid as long as $T\gg T_{co}$, and the limit
of the quasi-condensate regime, valid for $T\ll T_{co}$.   
Equivalently, in terms of chemical potential, as long as 
the chemical potential is positive and much larger than $\mu_{co}$ of 
Eq.~(\ref{eq.muco}), the gas is in the quasi-condensate regime 
whereas for negative chemical potential of absolute value much larger 
than $\mu_{co}$ the gas is in the ideal gas regime. 
This is illustrated 
in Fig.~\ref{fig.yynum}.
The two regimes differ by the fact that the $g^{(2)}(z)$ function 
is modified: it is close to one for any $z$ in the quasi-condensate 
regime while $g^{(2)}(0)=2$ in the ideal gas regime.

\subsubsection{Phase fluctuations}  
 In the quasi-condensate regime, although the gas is coherent with 
respect to the $g^{(2)}$ function, it is not coherent with respect to the 
$g^{(1)}$ function. This is why the gas is called a {\it quasi}-condensate. 
The phase fluctuations have been measured experimentally 
in various experiments where the quasi-condensate presented a one-dimensional
character~\cite{DetHelRyy01,HelCacKot03,RicGerThy03,ShvBugPet02,Ame08}. 
The description of the quasi-condensate given
above permits a simple calculation of those phase fluctuations 
as we now show.
Phase fluctuations are given by 
\begin{equation}
  \langle (\theta(z)-\theta(0))^2\rangle  
=\sum_{k>0} 2\langle \theta_{ck}^2\rangle (\cos(kz)-1)^2
+\sum_{k>0} 2\langle \theta_{sk}^2\rangle \sin^2(kz).
\end{equation}
Using Eq.~(\ref{eq.deltathetak}) and 
$(\cos(kz)-1)^2+\sin^2(kz)=2(1-\cos(kz))$  this gives
\begin{equation}
\langle (\theta(z)-\theta(0))^2\rangle = 4(mT/(Ln\hbar^2))
\sum_{k>0}\frac{1-\cos(kz)}{k^2}.
\end{equation}
Transforming $\sum_k$ into $L/(2\pi) \int_0^\infty dk$ and using
$\int_0^\infty (1-\cos(kz))/k^2 dk=\pi z/2$, we obtain
\begin{equation}
\langle (\theta(z)-\theta(0))^2\rangle = \frac{mTz}{n\hbar^2}
=\frac{2\pi z}{n\lambda_{dB}^2}.
\end{equation}
Since density fluctuations are very small, the $g^{(1)}$ function is
about $g^{(1)}(z)=n\langle e^{i(\theta(z)-\theta(0))}\rangle$.
Since the Hamiltonian is quadratic, we can use the Wick theorem
to compute  $\langle e^{i(\theta(z)-\theta(0))}\rangle$, which gives
 $\langle e^{i(\theta(z)-\theta(0))}\rangle=
e^{- \langle (\theta(z)-\theta(0))^2\rangle /2}$.
We find
\begin{equation}
g^{(1)}(z)\simeq e^{-mTz/(2n\hbar^2)}.
\label{eq.g1quasibec}
\end{equation}
 Comparing this to Eq.~(\ref{eq.g1iddegene}), we observe that the behavior of 
$g^{(1)}$ is close to that in the 
ideal gas regime. 
The factor of 2 difference in the correlation length formulae 
is because for the ideal gas regime, both density and phase 
fluctuations contribute to $g^{(1)}$ whereas only phase fluctuations remain
in the quasi-condensate regime.
 The crossover from the ideal gas regime to the quasi-condensate regime, 
at a temperature $T_{co}$, Eq.~(\ref{eq.frontncohom}), corresponds to 
the situation where the correlation length of phase fluctuations, given 
by Eq.~(\ref{eq.g1quasibec}), equals the correlation length 
of density fluctuations given by Eq.~(\ref{eq.xi}).

 In both this section and the previous one, we assumed that the temperature is 
high enough that the population of the relevant
modes (whose wavelengths are of the order of $\xi$) is much greater than unity. 
This is no longer the case when $T$ reaches values of the order 
or smaller than $gn$.  For lower temperatures, quantum fluctuations 
are expected to be dominant. This is the so-called quantum quasi-condensate
and the boundary between the thermal quasi-condensate regime and the 
quantum quasi-condensate regime is at $T\approx ng$, corresponding to
\begin{equation}
t\approx\frac{1}{\gamma}
\label{eq.qbecthqu}
\end{equation}
and is shown as a line in Fig.~\ref{fig.domaines}.
A recent experiment using an atom chip
observed these quantum 
phase fluctuations~\cite{Hofferberth:2008}. 
In the experiments we describe here however, the temperature is high enough 
that thermal fluctuations dominate.

\subsection{Exact thermodynamics}
\label{sec.exact}
In sections~\ref{sec.idealgas} and \ref{sec.quasibec}, we have discussed
models that apply independently in the asymptotic limits
of the nearly ideal gas regime ($T\gg T_{co}$ or equivalently 
$-\mu\gg \mu_{co}$) and
the quasi-condensate regime ($T\ll T_{co}$ or equivalently 
$\mu\gg \mu_{co}$) respectively.
While the above classification gives very useful insight, it should be
emphasized that the boundary between these two regimes is 
a  smooth crossover,
not a sharp transition and that neither of the two theories presented above 
account for the physics in the vicinity of the crossover.
Since in many cases we are interested
in the precise behavior near the crossover from 
the ideal gas to the quasi-condensate regime, it is not
sufficient to use the asymptotic results. 

As already mentioned in the introduction,  
the 1D Bose gas with repulsive delta-function interactions is
an example of an exactly solvable model  \cite{KorBogIze93,Tak99},
This allows
us to quantitatively compare predictions of the two approximate
descriptions to the exact results, and  verify the regions of
validity of the approximations. Furthermore the exact results
will turn out to be important for an accurate description of 
the experiments.

Exactly solvable models typically occur in lower dimensions
(1D quantum systems  \cite{KorBogIze93,Tak99} and 2D classical systems 
\cite{Bax82}) and 
allow one to obtain exact solutions for 
the quantum many-body eigenstates through a method known as the
``Bethe Ansatz'' (due to Hans Bethe \cite{Bet31}), for {\em any}
 value of the interaction strength.
For the repulsive delta-interacting 1D Bose gas (with periodic
boundary conditons), these solutions
were first obtained by Eliot Lieb and Werner Liniger
 \cite{LieLin63,Lie63}.
Furthermore,  the method based on the 
Bethe Ansatz can be extended to also obtain the thermodynamics exactly
(for {\em any} temperature),
 via a method 
due to C.~N. Yang and C.~P. Yang \cite{YanYan69}.

For a concise and lucid description of the Yang-Yang method to
obtain the exact thermodynamics of the 1D Bose gas and
the related equations, we refer the reader to the original literature
\cite{YanYan69}. In brief, each exact quantum many-body eigenstate
of the Lieb-Liniger hamiltonian Eq.~(\ref{eq.ham}) is characterized
by a set of distinct integer quantum numbers and a corresponding set of
distinct quasi-momenta $k$, obtained through the Bethe Ansatz. 
For a large system, one can consider the distribution 
of these quasi-momenta $\rho(k)$ and also of the ``holes''
$\rho_h(k)$, the latter corresponding to the ``missing'' values in 
the set of integers characterizing the individual quantum states.
By considering the entropy for given distributions $\rho(k)$ and
$\rho_h(k)$, Yang and Yang showed that the condition of thermal
equilibrium leads to a set of nonlinear integral equations that can be
solved by iteration. Subsequently, from the resulting distributions 
thermodynamic quantities such as pressure and free energy 
can be obtained.  Once these quantities 
have been found, further thermodynamic quantities can be calculated
using the standard thermodynamic relations.

Although numerical solutions to the Yang-Yang equations were
already obtained at an early stage  by C. P. Yang \cite{Yan70},
important further insight into the Yang-Yang thermodynamics
was gained much more recently by Kheruntsyan, Gangardt,
Drummond and Shlyapnikov \cite{KheGanDru03,KheGanDru05}. 
They 
calculated both density  and the 
normalized local density-density correlation
function $\gtwo$, and compared to approximate results in 
the various regimes discussed above.
The former, $n(\mu,T)$, is obtained as part of the equation of
state. The latter is obtained from the derivative of free energy
with respect to the coupling constant $g$, using the Hellmann-Feynman theorem.

\begin{figure}
\centerline{\includegraphics[width=9cm]{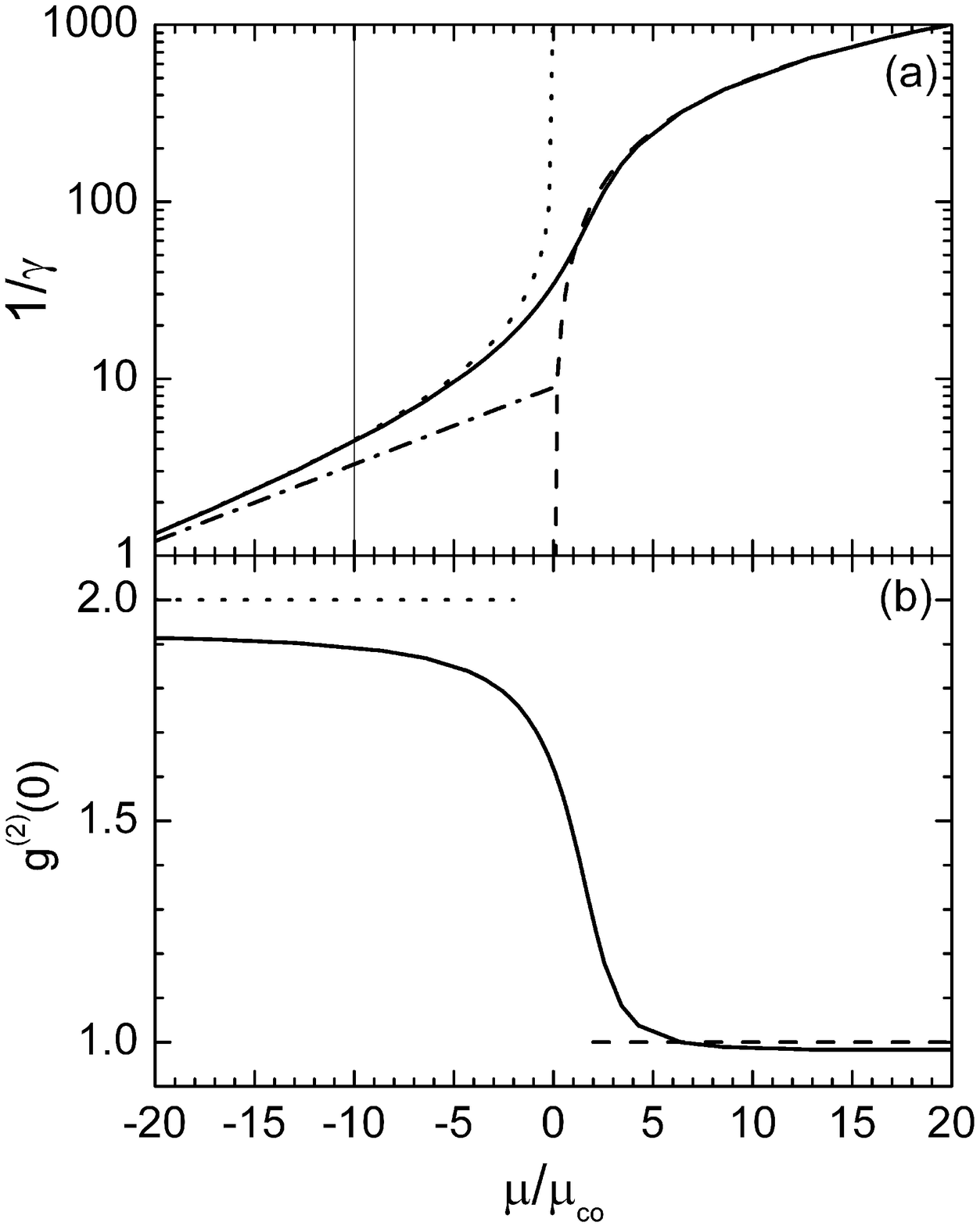}}
\caption{\it Normalised density ($1/\gamma$) and local pair correlation
$\gtwo$ as a function of chemical potential scaled to the crossover 
chemical potential $\mu_{co}$ given in Eq.~(\ref{eq.muco}) for fixed
temperature corresponding to $t=1000$.
Numerical results from the Yang-Yang equations (solid
lines, courtesy K. Kheruntsyan) are compared to the 
ideal Bose gas model (dotted line in (a), Eq.~(\ref{eq.stateid})) and 
the quasi-condensate model (dashed line in (a), Eq.~(\ref{eq.muqbec})).
The vertical line in (a) indicates the degeneracy chemical potential $-\mu=T$.
The classical Maxwell-Boltzmann prediction Eq.~(\ref{eq.stateMB})
is shown as dashed-dotted line. In (b) the asymptotic values of $\gtwo$ 
are indicated for both  the ideal-gas regime ($\gtwo=2$ for $\mu\ll -\mu_{co}$, dotted line), 
and  the quasi-condensate regime ($\gtwo=1$ for $\mu\gg \mu_{co}$,
 dashed line).  Adapted from Refs.~\cite{bouchoule:031606,Ame08}.
} 
\label{fig.yynum}
\end{figure}

As an important example, a comparison
to the approximate results of the previous sections is shown
in Fig.~\ref{fig.yynum}, for a fixed scaled temperature of $t=1000$. 
This value is in the relevant range for the experiments to
be described below.
Such curves as a function of chemical potential $\mu$ 
are particularly useful to describe
the behavior in a trap, since in this case one has a well-defined
global temperature, while the density varies (within the local
density approximation) with the local chemical potential $\mu(z)$
according to $\mu(z)=\mu-V(z)$, where $V(z)$ is the trapping 
potential. This will be discussed in more detail
in the following section.

Figure~\ref{fig.yynum}(a)  
shows that the exact density $(n\propto 1/\gamma)$  indeed
approaches the ideal-gas behavior as $\mu/\mu_{co}$ becomes sufficiently
negative, while for large positive $\mu/\mu_{co}$
it approaches the quasi-condensate result. 
 There is a large range in density (more than a factor 4) over 
which neither  asymptotic description gives correct predictions.
In the same vein, the  local density-density correlation
function $\gtwo$  (Fig.~\ref{fig.yynum}(b)) smoothly crosses over
from 2, the value for an ideal gas, to about 1, as expected for
a quasi-condensate. 
This smoothness is characteristic of
crossover behavior, and is drastically different from the step-like
behavior typical for a 3D gas. 

Looking more closely at Fig.~\ref{fig.yynum}, one sees that
the ideal gas description begins to fail 
for a gas that is only moderately 
degenerate: already at $\mu/T=-0.5$  
($\mu/\mu_{co}=-5$ for the considered $t$ parameter), 
a chemical potential for which  $n\lambda_{dB}\approx 10$ and
the population in the $k=0$ mode according to Eq.~(\ref{eq.bosedistr}) 
is $\approx 1.5$, the ideal Bose gas prediction 
is off by about 10\%. 
 This is because the  interaction-induced crossover
is sufficiently wide 
that for the 
used value of $t$ (1000), the chemical potential at degeneracy 
($\mu/T\approx -1$) is not very far removed.
 The narrowness of the degenerate ideal gas regime is also 
seen in Fig.~\ref{fig.domaines}.  To achieve  well separated 
regimes, one would need to work at much higher
$t$ and much smaller $\gamma$.
For $t=1000$, the effect of degeneracy is
nevertheless visible before the quasi-condensate crossover. 
This is shown by comparing the density with both 
the true ideal gas model and the  
Maxwell-Boltzmann model in Fig.~\ref{fig.yynum}: at $\mu/\mu_{co}\simeq -5$ 
the ideal gas model gives a prediction for the density accurate 
within 10\% as mentioned above (and the  nearly-ideal-gas description
can thus be expected be applicable) while the Maxwell-Boltzmann prediction 
is off by a factor of about 2.  

 Concerning the local pair correlation function $\gtwo$, it
deviates from the ideal-gas value of 2 for the entire range 
plotted in the figure. 
 The experiments presented in this chapter  however 
(see Sec. \ref{sec.experiments}), are not 
precise enough to detect this deviation. 
Finally, the 
value of $\gtwo$ can take values below unity in the 
quasi-condensate regime. We will briefly return to this
point in Sec.~\ref{sec.quasibecfluc}. 

Despite its power, the Yang-Yang theory 
does not permit calculation of any non-thermodynamic quantities. 
For example, only the {\em local} value of the density
correlation function $\gtwo$
has be obtained from thermodynamics, 
while the 
full behavior of $g^{(2)}(z)$ has been obtained from the exact solution 
only at zero temperature~\cite{caux:031605}.
 At finite temperature, the behavior of $g^{(2)}(z)$ 
has been obtained only by perturbative calculations valid 
in each asymptotic regime~\cite{sykes:160406}, but they do not describe 
the crossover itself.
  An alternative approach 
uses the fact that the crossover
appears in a highly degenerate gas. In this case, the modes are highly 
populated and a classical field approach is 
possible~\cite{Castin1Dclass,blakie:063608}.

\section{1D gases in the real world}
\label{sec.1Dtrap}
In real experimental situations, the atomic gas is neither homogeneous
nor purely one-dimensional. As usual, in our experiments the trapping is 
to a good approximation harmonic. The trap has cylindrical symmetry and
is characterized by a tight radial trapping frequency $\omega_\perp$
and a much lower axial trapping frequency $\omega$.
Here, we briefly summarize the main issues related to realizing
a 1D system in this trapping geometry.
 We first discuss the link between 
transverse effects related to $\omega_\perp$ and
we  present a model based on the  Yang-Yang thermodynamics, 
valid at low enough linear densities, 
that takes into account these  transverse degrees of freedom. 
We then discuss the effect 
of the longitudinal trapping potential. We finish 
by discussing the link with the 3D physics, in particular with regard
to the usual Bose-Einstein condensation in 3D.

\subsection{Transverse trapping and nearly 1D Bose gases}
Strictly speaking the conditions to be 1D in a transversely trapped gas
are that both temperature  and chemical potential
are much smaller than the radial vibration quantum,
 $T, \mu \ll \hbar\omega_\perp$.
 If this is the case, the gas is frozen in the transverse direction
both thermally and in terms of chemical potential, and the
 (many-body)
wave functions can be factorized into 
the product of a transverse part (the gaussian 
ground-state wavefunction of the radial trap)  and an axial part.
The system is then kinematically one-dimensional. 
Studying the scattering properties, 
Ref.~\cite{Ols98} has shown that the interactions can be modeled 
by an effective 1D coupling constant $g$ and, 
as long as the 3D scattering length $a$ is much smaller than the 
typical size of the transverse oscillator wavefunction,
$l_\perp=\sqrt{\hbar/m\omega_\perp}$, 
\begin{equation}
g=2 a \hbar\omega_\perp.
\label{eq.g1D}
\end{equation}

In most experiments on atom chips,
 neither of the above conditions on temperature
and chemical potential are well-fullfilled, and is it
necessary to also take into account the transverse degrees of freedom.

It is useful to consider the linear density \nlin ,
obtained from the actual 
3D density $\rho(x,y,z)$ through integration 
\begin{equation}
 \nlin(z)=\int\int dx dy \rho(x,y,z).
\end{equation}
When the gas is strictly 1D, one can identify $\nlin$ with the
1D density $n$. We will present our main
experimental results in terms of this linear density, 
because it turns out that  $\nlin$ 
is often the key parameter,
in particular when considering the crossover to 3D at
low temperatures.

This is in particular true for the quasi-condensate regime, at 
temperatures  $T\ll \hbar\omega_\perp$. In this regime, 
the chemical potential is close to its value at zero temperature, 
which is given by the solution of the radial 
Gross-Pitaevskii equation\cite{MenStr02,Ger04}.
It was found from comparison to numerical integration of the
radial Gross-Pitaevskii equation  \cite{MenStr02,Ger04}
that, in the quasi-condensate regime,
to  good approximation the chemical potential
can be expressed as\footnote{An additional factor $-1$ has
been introduced in brackets 
in Eq.~(\ref{eq.muGerbier}) compared to Ref.~\cite{Ger04}. This 
subtracts the radial zero-point energy 
$\hbar\omega_\perp$, 
so that  $\mu=0$ corresponds to the energy of the lowest energy ($k=0$) state,
 as in the treatment in Sec.~\ref{sec.theory}.}
\begin{equation}
\mu=\hbar\omega_\perp\left(\sqrt{1+4\nlin a}-1\right).
\label{eq.muGerbier}
\end{equation}
For linear density $\nlin\ll 1/4a$, we find that 
$\mu\simeq 2\hbar\omega_\perp a n$. We recover here the 
chemical potential $gn$ of the 1D case. At larger linear
density, the chemical potential is reduced compared to the formula 
$2\hbar\omega_\perp a \nlin$. This reflects the fact that, 
for large densities, the transverse cloud size is increased with respect 
to the transverse vibrational ground state.
As another example of how $\nlin$ is the relevant quantity for low
enough temperatures,
we note that the expression Eq.~(\ref{eq.g1quasibec}) for the 
phase coherence length
remains correct also on the 3D side of the crossover, if we replace
the 1D density $n$ by the linear density \nlin~\cite{PetShlWal01}.

\subsection{Applying 1D thermodynamics to a 3D trapped gas}
\label{sec.yy-model}
Another case that one can consider is when the interaction
energy is in the 1D regime,  $ng \ll \hbar\omega_\perp$, while
temperature is in the 1D-3D crossover, 
$T\simeq \hbar\omega_\perp$. 
A model for this regime was introduced in Ref.~\cite{AmeEsWic08},
and we describe it here.
The key step is to separately consider the radial states.
Under the above conditions only
the radial ground state is significantly affected by the interactions,
while the radially excited states can still be treated as an ideal gas.
Thus, for the radial ground state, the solution $n_{YY}(\mu,T)$ to the
Yang-Yang equations must be used.
Each radially excited state with radial quantum number $j\geq 1$ is now
considered as an independent ideal 1D gas, in thermal equilibrium
with the rest of the cloud. Each of the radially excited states
 is thus taken to have a density (cf. Eq.~(\ref{eq.stateid}))
\begin{equation}
\label{eq.nexc}
n_e(\mu_j,T) =\frac{1}{\lambda_{dB}}g_{1/2}(\exp(\mu_j/T)),
\end{equation}
where an effective chemical potential $\mu_j$ has been introduced that
takes into account the radial excitation energy,
\begin{equation}
\mu_j=\mu-j\hbar\omega_\perp.
\end{equation}
Taking into account the degeneracy factor $j+1$ of the radially excited
states, the total linear density in this model thus becomes
\begin{equation}
\nlin(\mu ,T)=n_{YY}(\mu ,T)+\sum_{j=1}^{\infty
}(j+1)n_{e}(\mu _{j},T). 
\end{equation}
As long as $\mu<\hbar\omega_\perp$, we have $\mu_j<0$ which
is necessary to avoid divergence of $g_{1/2}$ in Eq.~(\ref{eq.nexc}). 
In fact, from the
previous discussion in Sec.~\ref{sec.exact}, 
for our parameters ($t\approx 1000$), we can expect
an ideal-gas treatment of the radially excited density 
to begin to break down for $\mu_j/T>-0.5$  since
this is where interactions will 
become important. In practical cases where $T\approx\hbar\omega_\perp$,
the model should thus be accurate as long as $\mu<0.5\hbar\omega_\perp$,
while for $\mu>0.5\hbar\omega_\perp$ the model will start to become inaccurate.

\subsection{Longitudinal trapping}
 Experimentally, cold gases are axially confined in a confining potential $V(z)$ 
and the cloud is  
not infinite and homogeneous as assumed in the previous section. 
However, as seen below, for weak 
enough axial confinement, the results for homogeneous gases can be applied 
using a local density approximation. In the first following sub-section,
we present the local density approximation and 
discuss its predictions. 
We then evaluate the condition of validity of this approximation.

\subsubsection{Local density approximation}  
If the confinement is  
weak enough  that the correlation length of the gas is, at each position,
much smaller than the length of the mean density variations, then the 
gas may be divided into small slices in which the thermodynamics of uniform 
systems   
applies. A slice located at position $z$ 
is in equilibrium with the rest of 
the gas. It is thus described by the grand canonical ensemble 
at temperature $T$ and at a chemical potential $\mu_0$.
The energy of the gas  contained in this slice is shifted by 
the quantity $V(z)$. It is equivalent
to assuming that the chemical potential is $\mu_0 -V(z)$,
while the energy of the gas is unshifted.
 Thus, the local properties of the gas are that of a homogeneous infinite 
 gas at  temperature $T$ and local chemical potential 
$\mu(z)=\mu_0-V(z)$. This is the so-called local density approximation.

 Within the local density approximation, all the results
presented in the previous 
section hold. Thus, performing local analysis, 
one can  observe all the features of homogeneous
1D gases: the presence of the ideal gas regime, which includes the 
degenerate regime, the crossover towards a quasi-condensate and the 
quasi-condensate regime.  
 In particular, a quasi-condensate  appears in the 
center of the trap, when the peak density exceeds
 the crossover density $n_{co}$ given by Eq.~(\ref{eq.ncohom}).

 It is often  interesting 
to investigate the behavior of the gas using the extensive variable 
$N$, where $N$ is the total atom number. As long as the peak density 
is much smaller than $n_{co}$, the density profile is well described 
using the equation of state $n(\mu,T)$ of an ideal Bose gas. 
Then the total atom number is easily computed and, for gases 
that are degenerate at the trap center, we obtain~\cite{bouchoule:031606}
\begin{equation}
N=T/(\hbar\omega)\ln(T/|\mu_0|).
\label{eq.Ntrap}
\end{equation}
 The atom number at the 
crossover towards a quasi-condensate is obtained when the peak density 
reaches $n_{co}$. Inserting  Eq.~(\ref{eq.mudegenerate})
and Eq.~(\ref{eq.ncohom}) into  Eq.~(\ref{eq.Ntrap})
we find 
that the atom number at the crossover is approximately
\begin{equation}
N_{co}=T/(\hbar\omega)\ln\left (
(\hbar^2T/(mg^2))^{1/3}\right )=T/(3\hbar\omega)\ln (t/2).
\label{eq.Ncotrap}
\end{equation}
Since $t^{1/3}\gg 1$ (see text below Eq.~(\ref{eq.ncohom})), this equation 
can be inverted to give a crossover temperature 
\begin{equation}
T_{co}=N\hbar\omega/\ln\left (
(N\hbar^3\omega/(mg^2))^{1/3}\right ).
\end{equation}
 A comparison of this formula with a numerical calculation 
using Yang-Yang thermodynamics shows very good agreement \cite{bouchoule:031606}.
Since $t\gg 1$ at the crossover, Eq.~(\ref{eq.Ncotrap}) shows that the 
ratio $N_{co}/N_d$, where $N_d=\hbar\omega/T$ is 
the atom number at degeneracy, is larger than one at the crossover. 
Thus, even considering the extensive variable $N$, the 
 degenerate ideal gas regime is in principle identifiable.
However, the ratio $N_{co}/N_d$  only grows as logarithm of 
$t$ and it is in practice difficult to have $N_{co}/N_d$ very large.

\subsubsection{Validity of the local density approximation}
All the previous results use the local density approximation, which 
requires that the correlation length $l_c$ of the gas be much smaller 
than the scale $L$ of variation of the density. At the crossover, 
the correlation length of the gas is about $l_c\simeq\xi=\hbar/\sqrt{mgn}$, 
as seen in section~\ref{sec.theory}. 
To estimate $L$, let us approach the crossover from the ideal 
gas regime. The density profile of the central part of the cloud, obtained 
using~Eq.~(\ref{eq.mudegenerate})
and the local chemical potential $\mu(z)=\mu_0 -m\omega^2z^2/2$,
turns out to be a lorentzian of width $\sqrt{|\mu_0 |/m\omega^2}$. Thus, 
$L\simeq \sqrt{|\mu_0 |/m\omega^2}\simeq (gT_{co}/m\hbar\omega^3)^{1/3}$
at the crossover.
We thus find that the condition of validity of the local density approximation,
$l_c\ll L$, can be rewritten as
 \begin{equation}
 \omega\ll \omega_{co}=\frac{( E_g T^2 )^{1/3}}{\hbar}=\frac{\mu_{co}}{\hbar},
 \label{eq.condtrans}
 \end{equation} 
a  result which has been derived in~\cite{bouchoule:031606}.

 If the local density approximation (\ref{eq.condtrans}) is not satisfied, 
 the discrete structure of the trap energy levels has to be taken 
into account. In the opposite limit,  $\omega\gg \omega_{co}$, 
the quantization of energy levels plays a role while the gas 
is still described by an ideal Bose gas. Then, it has been 
shown in~\cite{Ketterlecond1D}
that one expects a condensation phenomenon to occur at a temperature
\begin{equation}
T_C=N\hbar\omega/\ln(2N).
\end{equation}
In contrast to the crossover described in the previous section (referred to
now as the interaction-induced crossover), 
this is a finite size phenomenon since $T$ goes to zero when the trap 
confinement $\omega$  goes to 0, $N\omega$ being fixed. 
This condensation phenomenon will dominate the 
interaction induced crossover when $T_{C}> T_{co}$. This condition 
is equivalent to $\omega \gg \omega_{co}$, which shows consistency 
of our analysis.

Experimentally, the condition (\ref{eq.condtrans}) to 
observe the interaction induced crossover is very easily 
satisfied:
using Eq.~(\ref{eq.g1D}),
the condition~(\ref{eq.condtrans}) reduces to
\begin{equation}
\omega\ll \omega_\perp (T/\hbar\omega_\perp)^{2/3} (a/l_\perp)^{2/3}.
\end{equation}
One can check that, for most alkali atoms, in trapping potentials 
with $\omega_\perp$ ranging from 1 to several tens of kilohertz and 
for temperatures between 0.1$\hbar\omega_\perp$ and $\hbar\omega_\perp$, 
this condition is easily fulfilled, unless $a$ is extremely 
small ($a<0.1$~nm). Thus, one expects that a trapped 1D gas undergoes 
the interaction induced crossover towards a quasi-condensate and that the 
local density approximation is valid to describe the gas.

\subsection{3D physics versus 1D physics}
 Experimentally, one expect a crossover from a one-dimensional behavior 
to a three dimensional behavior as the temperature of the gas increases
and, at large enough temperature, one expects to recover the physics of a 
three-dimensional gas.
 The physics of a 3D gas is very different from 
that of a 1D gas. The most striking difference is that, even in 
the absence of interactions, a 3D Bose gas  undergoes 
a phase transition towards a BEC due to saturation of the population 
of the excited states. This is in contrast to 1D gases where, in 
absence of interactions between atoms, the gas behaves, for any density, as 
a thermal gas in which bosonic bunching  is present.
For weakly interacting gases, in both 1D and 3D gases, a transition towards 
a (quasi-)condensate is expected. However, these transitions are different in 
nature and this difference can be captured by studying the validity 
of mean field theories in both cases.

 In 3D weakly interacting gases ($\rho a^3\ll 1$), the effect of 
interactions between atoms on the onset of Bose Einstein condenstation 
is very small. This is why 3D  Bose gases with weak interactions 
are well described 
by mean field theories. For instance, the thermodynamics is
given with a very good approximation by the 
Hartree-Fock-Bogoliubov self consistent 
theory~\cite{GriffinHF,PhysRevA.54.R4633}.
 In such a theory, at temperatures larger than the critical condensation 
temprature, the gas is described by the Hartree-Fock approach, in which 
correlations between atoms introduced by interactions are neglected. 
Condensation is then due, as for an ideal Bose gas, when the density 
reaches $2.612.../\lambda_{dB}$. For higher densities, a non zero 
condensate wave function appears, which is the order parameter of this 
second-order phase transition. 
 The experimental value of the critical temperature in weakly interacting 
ultra-cold Bose gases is in good agreeement with 
this theory~\cite{PhysRevLett.92.030405}. 
 
 However, even for weakly interacting gases, such a mean field theory 
is expected to fail very close to the critical point of temperature $T_c$. 
This is due to the 
large long wave length 
fluctuations that develop in the vicinity of the transition. 
In the condensate side, {\it i.e.} for $T< T_c$, the 
Hartree-Fock-Bogoliubov self consistent theory is valid only if 
the fluctuations of the condensate wave function, averaged over 
a volume of the order of the correlation length, are smaller than 
its mean-field value. This is the so-called Ginzburg criteria and it 
gives~\cite{PhysRevA.54.R4633}
\begin{equation}
\frac{T_c-T}{T_c}\gg a\rho^{1/3}.
\end{equation} 
The same criterion (up to an absolute value) is true above $T_c$.
The region around the transition where ${|T_c-T|}/{T_c}$ 
is 
of the order or smaller than  $a\rho^{1/3}$ is not expected 
to be described by a mean-field.  
Beyond mean-field effect include a modification 
of the transition temperature. Since interactions tend to decrease 
long wave length density fluctuations, they favor the appearance 
of a condensate and, for small paramater $a\rho^{1/3}$, an increase 
of the critical temperature is 
expected~\cite{PhysRevLett.83.2687,PhysRevLett.87.120401,PhysRevLett.87.120402}.
Such a modification is very small in cold atom experiments and has never 
been observed. 
 A second non mean-field effect is the modification of the critical 
exponent that describes the divergence of the correlation length 
in the vicinity of the critical point. Measuring beatnodes between 
the atomic field extracted at different places in the atomic cloud, 
the critical exponent was measured recently in dilute 
atomic gases, in agreement with beyond 
mean-field theories~\cite{Don07}. 

 The physics is very different in 1D systems, since long wavelength 
fluctuations  play an enhanced role compared to 3D systems. 
The crossover towards a quasi-condensate is, in 1D gases, a phenomenon 
driven by interactions. More precisely, the crossover towards a 
quasi-condensate is  produced by the correlations between atoms
brought by the interactions. 
 It cannot be captured by the Hartree-Fock theory 
because Hartree-Fock theory neglects correlations between atoms 
introduced by interactions.
Thus, in real systems 
the failure of the Hartree-Fock theory to describe the 
appearance of a (quasi-)condensate is a signature of the 1D nature of the 
physics involved.

\section{Experiments}
\label{sec.experiments}
 In this section we will discuss several experiments that have been carried 
out in both Orsay and Amsterdam using atom chips 
which probe the ideas discussed in the previous sections.
 Atom chip setups are very well suited to study one-dimensional geometry 
since very tight atom guides are easily realised by going close to a 
current-carrying micro-wire.
The atom chips which were used in the experiment presented below 
are sufficiently similar that
we will attempt to describe both at once. We will refer the reader to 
the individual experiments for more detailed information.
The atom chips we used employed current carrying wires to create magnetic trapping
fields for $^{87}$Rb atoms in the $F=2, m_F=2$ state. 
Magneto-optical traps, laser cooling and evaporative cooling were used
to load atoms into the chip-based traps, which tended to be 
highly confining but rather shallow.
Typical currents were on the order of a few amperes
 and the atoms were at a distance of several
tens of microns from the wire surface. 
Typical transverse confinement frequences ($\omega_\perp/2\pi$) 
were about 3 kHz,
while longitudinal frequencies were on the order of 10 Hz.
This transverse frequency corresponds to a temperature $\hbar\omega_\perp/k_B$
of 144~nK, and evaporative cooling 
was able to reach a temperature equal to or slightly above this value. 
For Rb atoms, with 3D scattering length $a=5.24$~nm,
the energy scale $E_g$ corresponds to 0.20~nK, and 144~nK
in reduced temperature units  corresponds to  $t=720$.
Since the longitudinal trapping potential is roughly harmonic, 
the linear atom density varied in space, and thus a single sample 
permits one to probe a large range in density at constant temperature. 
A single density profile thus corresponded to a horizontal line 
in Fig.~\ref{fig.domaines}.
The value of the parameter $\gamma$ was typically 
between $10^{-1}$ to $10^{-3}$.
The data consisted of absorption images of the cloud, taken
 either {\it in situ} or after a very short expansion time. 
Temperature measurements were made by fitting the wings of the cloud,
or by fitting to the Yang-Yang model (see description below). 

The first set of measurements we describe are simple observations of the
density profiles of nearly one-dimensional gases on an atom chip. 
The measurements were carried out with two purposes in mind. 
In the first measurements, carried out in Orsay, emphasis was placed
on proving that in the region of the crossover between the ideal gas 
and quasi-condensate regimes, no theoretical approach which neglected 
interaction induced correlations between particles, in particular the Hartree-Fock
approach, could explain the profiles.
In the second set, carried out in Amsterdam, it was shown that
the exact thermodynamic treatment accounted very well for the entire
observed profile, notably when the gas was in the crossover regime. 
After examining the profiles we move to another type of measurement
in which the absorption images were analyzed to give information about
density fluctuations. 
Although these measurements where chronologically the first,
we will treat them last. 

\subsection{Failure of the Hartree-Fock model}
\label{sec.HFfailure}
\begin{figure}
\centerline{\includegraphics{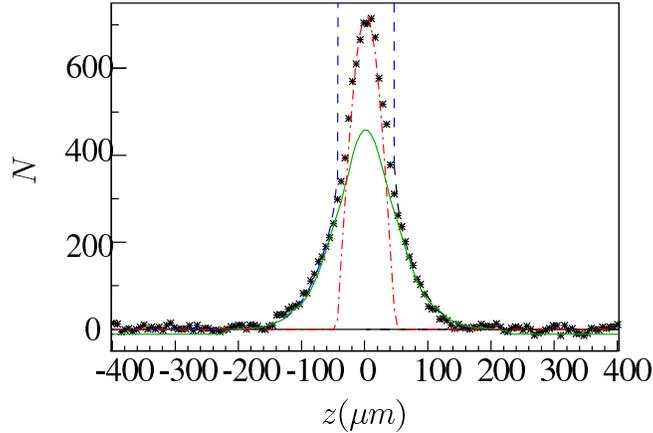}}
\caption{\it Failure of Hartree-Fock theory in a quasi-1D gas. 
The experimental profile (crosses) is compared with the profiles
expected for a quasi-condensate (dotted-dashed),
for an ideal Bose gas (dashed), 
and to the profile predicted by the Hartree-Fock theory (continuous line)
for the same temperature and chemical potential.
The vertical axis is the number of detected atoms per 6 $\mu$m longitudinal pixel.
The temperature of the gas was $T=360$~nK$=2.75\hbar\omega_\perp$. 
Adapted from Ref.~\cite{TreEstWes06}.
}
\label{fig.HF}
\end{figure}

A typical density profile is shown in Fig.~\ref{fig.HF}. 
Superimposed on the data are three different
theoretical predictions. 
The dashed line shows the profile as predicted
by the ideal gas model. 
In the wings of the profile this model should be valid, and 
indeed the fit to the wings of the distribution was used
to deduce the temperature and the chemical potential of the gas.
Clearly, however the ideal gas prediction begins to rapidly
deviate from the data, because, without interactions, a 1D Bose gas
can accommodate arbitrarily high densities at a given temperature. 
The dash-dotted line shows the prediction of the
quasi-condensate model Eq.~(\ref{eq.muGerbier}),
 at the same chemical potential as
was found by fitting the wings. 
This model accurately reproduces the high density part of the
distribution, but not the presence of so many atoms in the wings of the distribution.

The Hartree-Fock theory is a variational method in which the atoms are described 
by a gas of non interacting bosons subject to an effective potential 
$V_{HF}$ due to the mean field of the other atoms. 
Minimizing the free-energy of the gas, one finds 
\begin{equation}
V_{HF}({\bf r}) =2g_{3D}{\bf \rho}
\end{equation}
where $g_{3D}=4\pi\hbar^2a/m$ is the 3D coupling constant and $\rho$ 
is the 3D gas density. This theory is thus self consistent, since
for a given chemical potential and temperature, ${\bf \rho}$ 
depends on $V_{HF}$. The factor 2 reflects the bunching, 
which is present in the Hartree-Fock approximation since the gas 
is descibed by a gas of non interacting bosons.

Using minimisation techniques, 
the Hartree-Fock density 
profile was calculated in Ref.~\cite{TreEstWes06}
for the experimental three-dimensional 
trapping potential
and for the temperature and chemical potential found by fitting the wings of the distribution.
One sees that the Hartree-Fock density profile,
shown as a solid line in Fig.~\ref{fig.HF}, reproduces the wings of the
density profile, and does not diverge as does the ideal gas profile.
It does not however, reproduce the high density part of the profile.  
Moreover,  the Hartree-Fock calculation shows that the 
Hartree-Fock gas is far from being saturated:  the population 
of the ground state is very small, and no condensation
is expected according to this mean-field model. 
The excess of atoms in the center is the onset of 
a quasi-condensate, although the cloud is not deep into the quasi-condensate regime.
This peak is formed by the effects of interactions altering
the two body correlation function so as to lower the interaction energy relative
to a Hartree-Fock gas at the same density.

\subsection{Yang-Yang analysis}
\begin{figure}
\centerline{\includegraphics{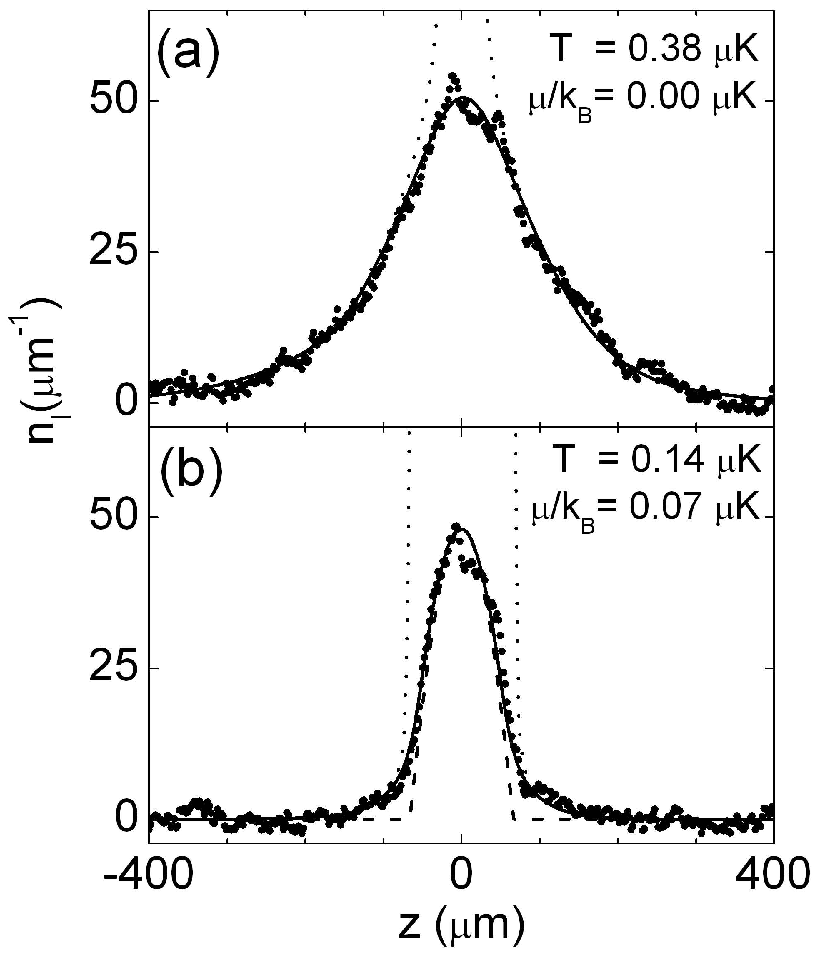}}
\caption{\it Comparison of experiment to Yang-Yang thermodynamics. 
 The model described in  Sec.~\ref{sec.yy-model} is
 fit to two examples of measured
 linear densities $\nlin$ (dots) in the Amsterdam experiment \cite{AmeEsWic08,Ame08}.
 The resulting fits (continuous curves)
 yield chemical potential $\mu$ and temperature $T$ as indicated.
Dotted curves: ideal-gas profile at the same temperature and chemical
potential exhibiting divergence for $\mu(z)=0$.
Dashed curve in (b): quasi-condensate profile with the same peak density 
as the experimental data.
In these experiments $\hbar\omega_\perp=158$~nK.
 Adapted from Ref.~\cite{AmeEsWic08}.
}
\label{fig.yy-exp}
\end{figure}

Two more examples of axial density profiles measured  \cite{AmeEsWic08} 
at two different temperatures
and a peak linear density of $\approx 50$~$\mu$m$^{-1}$
are shown in Fig.~\ref{fig.yy-exp}.
These profiles were fit to the model based on the
exact Yang-Yang solutions described in Sec.~\ref{sec.yy-model}.
The fits are shown in the Figure as continuous curves,
and the resulting temperature $T$ and chemical potential $\mu$ are also
indicated. The chemical potential $\mu$ and the temperature $T$ are the only
free parameters in the model, and it was found that the full
set of {\em in situ} 
measurements could be explained by the Yang-Yang-based model
\cite{AmeEsWic08}.  For comparison, The ideal-gas prediction and
the quasi-condensate prediction are also shown.
Clearly, the  Yang-Yang-based model describes the entire profiles
well, while the approximate models fail, in particular the smooth
crossover between the two approximate models in the region where
 $\mu(z)\approx 0$ is captured very well by the model.

This analysis was further corroborated by measurements of the
axial momentum distribution \cite{AmeEsWic08}, 
obtained using Bose gas focusing \cite{ShvBugPet02}.
The tails of the momentum distribution were used to extract
temperatures, and these were found to agree very well with the 
temperatures derived from the Yang-Yang fit to the {\em in situ}
data. The full momentum distribution
is not a thermodynamic quantity, and can thus not be obtained
directly from the Yang-Yang analysis.

The similarity of the measured density profiles of
Figures \ref{fig.yy-exp}(b) and \ref{fig.HF} clearly suggests that the
same physics of an interaction-induced crossover applies to both
experiments. Although it is tempting to apply the Yang-Yang-based analysis
of Sec.~\ref{sec.yy-model} also to the data of Fig.~\ref{fig.HF}, this has
not been done. It is likely that the result would not
be quantitatively accurate, because at the higher linear densities and temperatures
of Fig.~\ref{fig.HF}, the validity limits of the model of Sec.~\ref{sec.yy-model}
are reached near the peak of the profile
 (since both $\mu\approx \hbar\omega_\perp$ and $T\approx\hbar\omega_\perp$).
Interactions are then expected to also play a role
in the radially excited states, and also interactions among the different
radial states will be significant.

\subsection{Measurements of density fluctuations}
As we have emphasized in Sec.~\ref{sec.theory}, the transition towards a 
quasi-condensate in 1D gases is characterized by the inhibition of 
atom bunching, the large density fluctuations characteristic of a thermal Bose gas. 
A direct measurement of the density fluctuations 
through the crossover thus captures an essential characteristic of the
crossover.

The measurement of density fluctuations proceeds similarly to 
the density profile measurements.
The difference is that many (about 300) profiles are acquired and, roughly speaking,
for each observation pixel, we compute the variance of the density measurements as
well as the mean.
We can relate this variance to the density fluctuations predicted by various theoretical
approaches as described below.

The measurements involve several subtleties requiring careful normalizations
and corrections of the data. These are described in detail  
in~\cite{EstTreSch06,Tre07}.
The measurement requires a high degree of reproducibility in the data.
The atom chip geometry permits the construction of a very compact 
apparatus with low sensitivity to vibration. 
The images contain not only noise due to atom fluctuations, but also photon shot noise.
The photon noise must be carefully characterized and subtracted.
Examples of the data are shown in Figs.~\ref{fig.bunching} and \ref{fig.fluctuquasibec}.

\subsubsection{A local density analysis}
 The pixel size in the experiment is $\Delta=6~\mu$m. The pixel size is much larger 
than the  correlation length of the gas which is always smaller than a micron 
in these experiments, but much smaller 
than the longitudinal length scale of mean density variation. 
Thus, the data should reproduce number fluctuations predicted in a
longitudinal local density treatment. 
More precisely, 
the gas contained in the pixel located at position $z$ can be 
described as a gas, confined transversely by the transverse potential 
of frequency $\omega_\perp$ and confined longitudinally 
by a box like potential of  size $\Delta$. The properties of this slice, 
which can exchange 
energy and particles with the rest of the gas, is well described 
within the grand-canonical ensemble. 
 The energy shift $V(z)$ of 
this slice can be converted to a shift $-V(z)$ of the 
chemical potential. This is the local density 
approximation, already discussed in sec.~\ref{sec.1Dtrap}.
Since $\Delta$
is large compared to correlation length of the gas, the 
boundary conditions used to compute thermodynamic quantities are all equivalent 
and we use the periodic boundary conditions in the following.

\begin{figure}
\centerline{\includegraphics{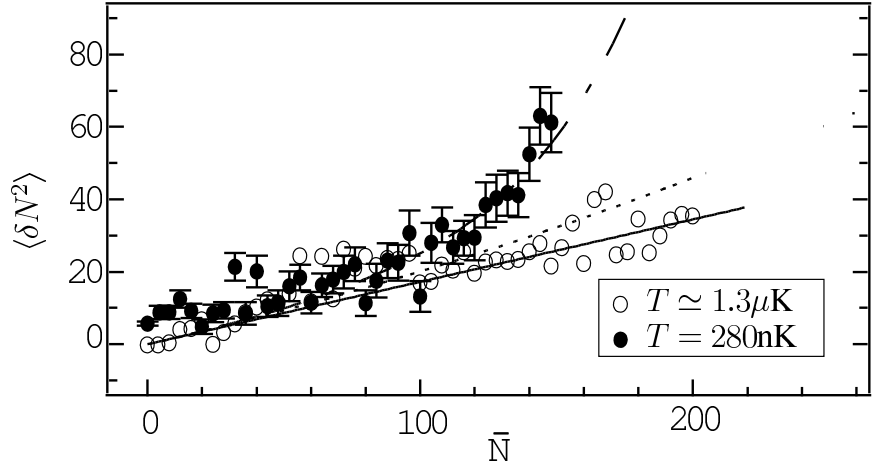}}
\caption{\it
Density fluctuations of a gas on an atom chip.
The atom number variance in an observation pixel (6 $\mu$m) is 
plotted as a function of the mean number. 
The open cirles are the fluctuations measured for a hot cloud 
($T=1.3~\mu$K corresponds to $10\hbar\omega_\perp$)
for which bunching is unobservable because of the large number of
transverse states involved. The variance is  due to atom shot noise. 
Full circles correspond to a colder cloud, at a temperature 
$T=2.1\hbar\omega_\perp$. The increase in fluctuations 
is due to bunching. 
The theoretical prediction for 
an ideal Bose gas at the same temperature is given by the dashed curve. 
The prediction for a nondegenerate cloud, Eq.~(\ref{eq.fluctunondegene}), 
is shown as the dotted curve.  The degeneracy of the gas is evident.
Adapted from Ref.~\cite{EstTreSch06}.
}
\label{fig.bunching}
\end{figure}

\begin{figure}
\centerline{\includegraphics{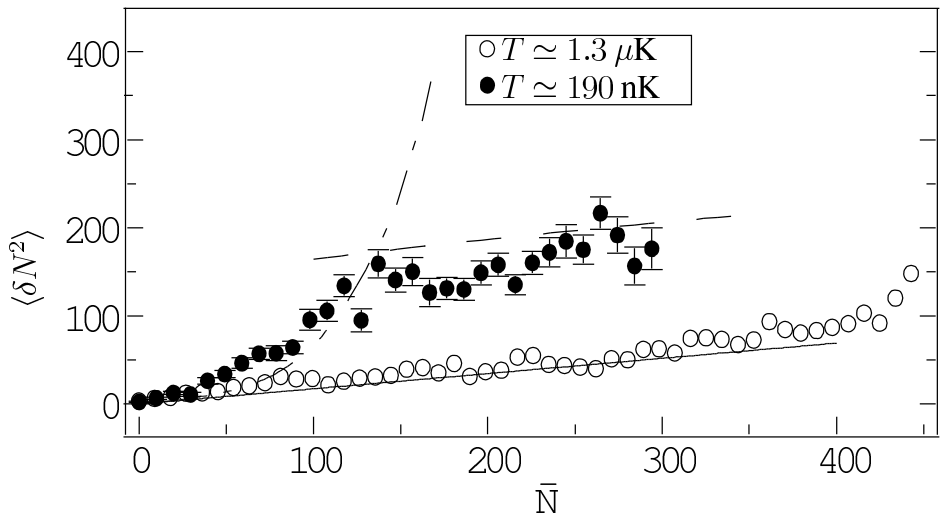}}
\caption{\it 
Density fluctuations in the quasi-condensate regime.
The dashed-dotted curve is the prediction for an ideal Bose 
gas at the same temperature as in figure \ref{fig.bunching}. The dashed curve is the prediction 
for a quasi-condensate. In units of transverse energy, the temperature is 
$T=1.4 \hbar\omega_\perp$.
Adapted from Ref.~\cite{EstTreSch06}.
}
\label{fig.fluctuquasibec}
\end{figure}

 Within the local density approximation, the 
confinement potential $V(z)$ is irrelevant to analyze 
the atom-number fluctuations. The atom number fluctuation $\delta N^2$ 
in each pixel depends only on the temperature $T$ and on 
the local chemical potential. Equivalently, $\delta N^2$ is 
a function of $T$ and $\langle N\rangle $, since the linear 
density is a monotonically increasing function of the chemical potential.
We thus choose, for each cloud temperature, 
to represent the measured atom number fluctuation as a function
of the mean atom number in the pixel. Experimental results are shown 
in Fig.~\ref{fig.bunching} and fig.~\ref{fig.fluctuquasibec}.

\subsubsection{Ideal gas regime : observation of bunching}
If the gas within a pixel can be considered ideal, we can use the results of 
Sec.~\ref{sec.idealgas}.
The fluctuations of atom number $n_i$ in each one-atom quantum state $|i\rangle$
are 
\begin{equation}
\langle n_i^2\rangle -\langle n_i\rangle^2=\langle n_i\rangle
+\langle n_i\rangle^2.
\end{equation}
The fluctuations of the total atom number $N$ are thus
\begin{equation}
\langle N^2\rangle -\langle N\rangle ^2=\langle N\rangle+
\sum_i \langle n_i\rangle ^2,
\label{eq.bunchingsumetats}
\end{equation}
where the sum is performed over all the quantum states. 
The mean values $\langle n_i\rangle$ are given by the 
Bose distribution and the fluctuations of $N$ are easily 
computed. 

A rough calculation is as follows: if $M$ quantum states are populated with similar populations, 
Eq.~(\ref{eq.bunchingsumetats}) simplifies to 
\begin{equation}
\langle N^2\rangle -\langle N\rangle ^2=\langle N\rangle 
+\langle N\rangle \frac{\langle N\rangle}{M}.
\label{eq.bunchingM}
\end{equation}
The first term of the right hand side is the shot noise term, expected
for uncorrelated, statistically independent atoms. 
The second term on the right hand side is 
the effect of the  bunching. We see from this expression that 
as long as $\langle N\rangle /M$ is much smaller than 1, the bunching term 
is negligible 
compared to the shot noise term. The ratio $\langle N\rangle /M$ is 
approximately the phase space density of the gas and is much smaller 
than 1 if  the gas is non degenerate. 
 Thus one expects  the measured atom number fluctuations 
to be dominated by the shot noise term for gases at high temperature. 
This is observed experimentally for non-degenerate clouds, 
as shown by the open circles in Fig.~\ref{fig.bunching}.
The linearity of the measured value of 
$\langle N^2\rangle -\langle N\rangle ^2$ versus $\langle N\rangle$
shows that the fluctuations are given by the shot noise. The fact that 
the slope is smaller than the expected slope of 1 is due to the fact 
that the optical resolution (about 10~$\mu$m) is larger
 than the pixel size \cite{EstTreSch06}. 

One can also give a more precise calculation of the fluctuations.
For this purpose, we index the quantum states 
by the integer $n_x$ and $n_y$, which 
label the transverse vibrational levels, and the longitudinal 
wave vector $k_z$, which takes values in multiples of $2\pi/\Delta$.
For a highly nondegenerate gas, $|\mu|\gg T$, the population of each state is 
given by the Boltzmann law
\begin{equation}
\langle n_{n_x,n_y,k_z}\rangle =Ae^{-(\hbar^2k_z^2/2m +\hbar\omega_\perp(n_x+n_y))/T}, 
\end{equation}
where the normalization factor $A$ is 
\begin{equation}
A=\frac{N\lambda_{dB}}{\Delta \sqrt{2\pi}}
\left (1-e^{-\hbar\omega_\perp/T}\right )^2.
\end{equation} 
Inserting this into Eq.~(\ref{eq.bunchingsumetats}), we obtain 
\begin{equation}
\langle N^2\rangle -\langle N\rangle ^2=\langle N\rangle
+ \langle N\rangle ^2 \frac{\lambda_{dB}}{\sqrt{2}\Delta}\tanh^2(\hbar\omega_\perp/2T).
\label{eq.fluctunondegene}
\end{equation}
 We thus recover an expression similar to Eq.~(\ref{eq.bunchingM}), with 
$M={\sqrt{2}\Delta}/(\lambda_{dB}\tanh^2(\hbar\omega_\perp/2T))$. 
The $\tanh$ term accounts for the number of populated transverse states. 
The term ${\sqrt{2}\Delta}/\lambda_{dB}$, which accounts for the 
longitudinal states, may be recovered by a semiclassical analysis:
the volume of the occupied phase space is $\Omega\simeq \Delta \sqrt{mT}$
and the number of quantum states contained in this volume is of the 
order of $\Omega/\hbar$.  
Equation~(\ref{eq.fluctunondegene}) is valid if the gas is  
non degenerate. 
When the gas becomes degenerate, the distribution 
of the mean occupation number $\langle n\rangle$ versus the state 
energy becomes more 
peaked around zero. This amounts to a reduction of the effective number of
occupied states $M$ and the effect of bunching is larger than the 
prediction of Eq.~(\ref{eq.fluctunondegene}). 
For highly degenerate gases, Eq.~(\ref{eq.fluctunondegene})
underestimates  the true fluctuations, which become large 
compared to the shot noise level. 
 
The bunching effect is quite clear for a cold enough cloud 
as shown in Fig.~\ref{fig.bunching}. 
In this experiment, the bunching term is even larger than the shot noise term, 
indicating that the gas is degenerate. 
The degeneracy is also shown by a comparison of the data 
with Eq.~(\ref{eq.fluctunondegene}) shown as a dotted line.
This equation, valid for 
a non degenerate gas, underestimates the measured 
fluctuations. On the other hand, a calculation of 
Eq.~(\ref{eq.bunchingsumetats}) using the true Bose occupation
factor is in much better agreement with the data. 
This comparison shows that, at least as concerns fluctuations,
the gas is well described by an ideal, degenerate Bose gas.

\subsubsection{Quasi-condensate regime: saturation of atom number fluctuations}
\label{sec.quasibecfluc} 
At sufficiently high density and low temperature, repulsive interactions 
between atoms are no longer negligible. As described in 
section~\ref{sec.theory}, one expects the interactions to reduce 
the density fluctuations to lower the interaction energy. 
The gas then enters 
the quasi-condensate regime. 
For the temperature $T=2.1\,\hbar\omega_\perp$ of 
the data in Fig.~\ref{fig.bunching},
using Eq.~(\ref{eq.g1D}) and assuming a purely 1D gas, Eq.~(\ref{eq.ncohom})
gives a density at the crossover of about 130 atoms per pixel.
Although Eq.~(\ref{eq.ncohom}) does not apply since the gas is not purely 1D,
this rough estimate shows that the crossover to a quasi-condensate is achievable 
at slightly higher atom number and/or lower temperature. 
Measurements of atom number fluctuations 
in a regime where the cloud center is in the quasi-condensate regime 
are shown in Fig.~\ref{fig.fluctuquasibec}. Whereas at low atomic density 
the measured fluctuations are in agreement with the ideal Bose gas prediction, 
one sees a saturation of the fluctuations at higher densities. 

To calculate the fluctuations, we first suppose the gas to be purely one dimensional, 
with a coupling
constant $g$ given by Eq.~(\ref{eq.g1D}).  
 In a local density approximation, we consider 
atom number fluctuations in a longitudinal box of length $\Delta$
in equilibrium with a reservoir of energy at temperature $T$ and
a reservoir of particles at chemical potential $\mu$. 
 As explained in sec.~\ref{sec.quasibec}, the Hamiltonian is 
quadratic in $\delta n$ in the quasi-condensate approximation and 
$\delta n$ can be expanded as a sum of independent modes indexed by the wave vector $k$. 
The atom number fluctuations $N-\langle N\rangle$ 
are obtained by integrating $\delta n$ 
over the pixel size.
Thus, the  only excitation that leads to atom number fluctuation 
is the zero momentum mode. Its energy, derived from Eq.~(\ref{eq.Hphonons}),
is 
\begin{equation}
H_{k=0}=\Delta \frac{g}{2}\delta n_0^2.
\label{eq.Hk0}
\end{equation} 
Using the equipartition theorem, we find $\delta n_0^2=T/(g\Delta)$. 
The atom number fluctuations, which are  $(\Delta \delta n_0)^2$, are thus 
given by: 
\begin{equation}
\langle N^2\rangle -\langle N\rangle^2=\frac{\Delta T}{g}.
\label{eq.deltaToverg}
\end{equation}
Thus, we expect the atom number fluctuations to be independent of 
$\langle N \rangle$. 
The shot noise term is not present 
in this quasi-condensate regime: interactions between atoms 
prevent even the shot noise fluctuations and at temperature smaller 
than $gn$, one expects to observe sub-shotnoise fluctuations.
This feature is also seen in Fig.~\ref{fig.yynum} where $\gtwo$ 
goes below unity in the exact solution.

 In the experimental results shown in Fig.~\ref{fig.fluctuquasibec}, 
the typical interaction energy is about $0.7\hbar\omega_\perp$. 
In these conditions, the transverse degrees of freedom cannot 
be neglected and the result  of Eq.~(\ref{eq.deltaToverg}) must be corrected. 
More precisely, the phonons, which are longitudinal
density waves, are associated with a breathing of the transverse shape
of the cloud. 
For phonons of frequency much smaller than the transverse frequency, 
the transverse shape of the cloud follows adiabatically the 
ground state equilibrium state for the local linear density. 
We denote by $E_{eq}(n)$ the energy of the gas 
per unit length for a linear density $n$.
The phonon Hamiltonian of Eq.~(\ref{eq.Hphonons}) is the term of 
the Hamiltonian of order two in $\delta n$ and 
$\nabla \theta$. Thus,  
the interaction term of the phonons  is  
$
E_{\rm{int}}=\frac{1}{2} L 
(\partial^2 E_{eq}/\partial n^2)
\delta n_k^2$.
Since $\partial E_{eq}/\partial n$ is the chemical potential of the gas (to 
zero order in $\delta n$), we can rewrite the former expression as
 $E_{\rm{int}}=\frac{1}{2} L 
(\partial \mu/\partial n)
\delta n_k^2$.
In particular, the zero momentum term is 
\begin{equation}
H_{k=0}=\frac{1}{2} L 
\frac{\partial \mu}{\partial n}
\delta n_0^2.
\end{equation}
Then, the atom number fluctuations are
\begin{equation}
\langle N^2\rangle -\langle N\rangle^2=
\frac{\Delta T}{\partial \mu/\partial n}.
\label{eq.thermo}
\end{equation}
 Although we derived this expression  
in the approximate quasi-condensate theory 
using expansion of the Hamiltonian to second order in $\delta n$, 
we recover here a well known result  of statistical physics.
More precisely, as shown in~\cite{Kittel}, 
Eq.~(\ref{eq.thermo}) holds for any system in equilibrium 
with a particle reservoir 
at chemical potential $\mu$ and with an energy reservoir of temperature $T$.

To apply Eq.~(\ref{eq.thermo}) to the experiment, we need the 
equation of state $\mu(n)$.
Using Eq.~(\ref{eq.muGerbier}) and Eq.~(\ref{eq.thermo}), one can compute the expected 
fluctuations. Fig.~\ref{fig.fluctuquasibec} shows that the results 
are in fairly good agreement with the measured atom number fluctuations.

\section{Conclusion}
We hope that we have given the reader a useful overview of the physics of 1D gases in the
weakly interacting regime. 
These systems are rich and manifest several different regimes
separated by smooth crossovers. 
It is often necessary to appeal
to many different physical models to understand them. 
The existence of exact solutions allows us to test the models
and to explore their validity in the face of highly non-trivial many body correlations.
Although the experimental and theoretical work we have described is
quite extensive, we believe that much work remains to be done.
The study of fluctuation phenomena is still at an early stage. 
For example the exact thermodynamics should admit a careful comparison with 
data such as that in Fig.~\ref{fig.fluctuquasibec},
and improved experiments should probe larger parameter ranges and possibly
even permit measurements of the correlation length.
It may also be possible in the near future to enter the strongly interacting regime
using an atom chip.
Experiments are also capable of measuring momentum distributions
\cite{AmeEsWic08},
but so far no quantitative theoretical comparison has been made. 
Finally measurements of fluctuations and correlations in momentum space are
also experimentally feasible  \cite{HBTHe}.

\section{Acknowledgments}
We thank K. Kheruntsyan for a critical reading of the manuscript.
NJvD acknowledges stimulating discussions with 
J.T.M. Walraven, G.V. Shlyapnikov, J.S. Caux, and A.H. van Amerongen.
The Amsterdam work is supported by
FOM (Stichting Fundamenteel Onderzoek der Materie) and
NWO (Nederlandse Organisatie voor Wetenschappelijk Onderzoek).
The work in Orsay was supported by the Institut Francilien pour la Recherche 
en Atomes Froids, and by the SCALA program of the E.U.

\setlength{\bibindent}{4mm} 

\bibliography{fluctudens}

\end{document}